# Enhancing Phishing Email Identification with Large Language Models

Catherine Lee
catherine.lee@gatech.edu
Georgia Institute of Technology

*Abstract*—Phishing has long been a common tactic used by cybercriminals and continues to pose a significant threat in today's digital world. When phishing attacks become more advanced and sophisticated, there is an increasing need for effective methods to detect and prevent them. To address the challenging problem of detecting phishing emails, researchers have developed numerous solutions, in particular those based on machine learning (ML) algorithms. In this work, we take steps to study the efficacy of large language models (LLMs) in detecting phishing emails. The experiments show that the LLM achieves a high accuracy rate at high precision; importantly, it also provides interpretable evidence for the decisions.

## I. Introduction

Phishing is a prevalent form of cybercrime that involves the use of deceptive emails to trick recipients into divulging sensitive information, such as passwords, credit card numbers, and other personal data [16]. Modern phishing schemes are highly sophisticated, employing advanced social engineering tactics to increase their effectiveness [25]. In a report published by the Anti-Phishing Working Group (APWG) in 2024, it was noted that there were 116,473 unique phishing email campaigns observed in the first quarter of 2024 [1].

Researchers have made a concerted effort to develop effective solutions for the automatic detection of these threats. To fortify defenses against phishing emails, numerous studies have focused on machine learning (ML) and deep learning (DL) algorithms that facilitate the analysis and classification of emails [2, 11, 18, 19].

Current approaches in ML-based phishing detection can be categorized by the feature sets utilized with anti-phishing techniques: Sender Information, URL Analysis, Email Content Analysis, Header Information, and Stylometric Features [3, 19, 39, 24].

Hybrid feature selection often involves combining multiple feature selection methods to improve the accuracy and efficiency of identifying phishing emails [2, 26].

However, ML/DL models can overfit the training data, performing well on known examples but poorly on new, unseen phishing attempts, and lack the ability to learn and adapt to evolving attack patterns. As a result, they require frequent retraining to keep up with changes in data [11].

Large Language Models (LLMs), such as OpenAI's GPT-4 [23], are advanced DL models trained on vast amounts of text data, and LLMs can perform tasks without dedicated prior training [17]. These models leverage the transformer architecture to understand and generate human-like text, making them highly effective in natural language processing (NLP) tasks [22]. The capabilities of LLMs extend beyond simple text generation; they can analyze and interpret complex language patterns, detect anomalies, and identify subtle cues indicative of phishing attempts. One of the key advantages of LLMs in phishing detection is their ability to understand context and semantics [17]. LLMs can analyze the content of emails holistically, considering factors such as tone, intent, and linguistic nuances [27]. This enables LLMs to detect phishing emails that may bypass conventional filters by mimicking legitimate communication styles.

In this study, we propose using LLMs to detect phishing emails through hybrid feature selection. Utilizing a diverse dataset of phishing emails, we include a mixture of datasets featuring different types of phishing (such as spear phishing, traditional phishing, and generative AI (GenAI) phishing). With designed prompt engineering, LLMs leverage their powerful analytical capabilities to detect phishing emails, providing detailed explanations for their classification decisions. We will evaluate and compare the performance of four state-of-the-art LLMs with detection accuracy on their capability for phishing detection. Importantly, we further analyze the result generated by the LLMs on subsequent decisions on false positives and define the current gap in LLM phishing detection.

The structure of the paper is as follows: Section II provides an overview of the background of email phishing detection research of machine learning algorithms to classify emails based on the selected features and current studies and gaps on LLMs in cybersecurity and phishing email. Section III outlines the methodology and the decision-making feature selection. Section IV depicts the experimentation results, and Section V discusses the outcomes of the experimentation and indicates the future directions for this research. Section VI concludes the paper.

## II. Literature Review (Related Work)

Before utilizing LLM on phishing detection, many research papers have been published so far on the phishing detection problem using ML and DL approaches on analyzing the email's structure, such as headers information, URLs, syntax, attachments, etc. or on the email's body content [2, 4, 13].

Earlier works have primarily focused on URL-based feature analysis for phishing detection by analyzing its characteristics and patterns using ML [40, 41]. However, relying solely on URL-based features for phishing detection has its limitations. URLs can often be long and complicated, making them difficult to analyze briefly. Spammers take advantage of URL shorteners (tools like TinyURL) [4] to create shorter, more manageable URLs. These shortened URLs can be included in emails to hide the actual destination of the link. Phishers can obfuscate (hide) the true nature of URLs using these tools, making the links appear legitimate

[12, 29]. This allows them to bypass URL-based detection methods that might otherwise flag the link as suspicious.

Recent phishing email detection focuses on email content analysis applying NLP techniques, such as TF-IDF Word2Vec, and Word Embedding [5] to train classification machine learning algorithms from both phishing and legitimate emails to attain classifier model email classification.

Harikrishnan et al. [8] utilized TF-IDF (Term Frequency-Inverse Document Frequency) for preprocessing data, SVD/NMF for feature extraction, and dimensionality reduction to train classical machine learning techniques like Decision Tree and Random Forest. They train the model on two sub-tasks with the dataset that includes emails with email bodies and headers (sub-task 1) and emails with only bodies (sub-task 2). They achieved a high accuracy of 99.9% on RF. However, the authors mentioned that their ML classifiers overfit the testing data due to the imbalanced dataset.

A combination of the TF-IDF, Word2Vec, and BERT techniques with MLs was proposed in [5]. The authors utilized three state-of-the-art methods (i.e., TF-IDF, Word2Vec, and BERT) and five well-known ML algorithms (Logistic Regression, Decision Tree, Random Forest, Gradient Boosting Trees, and Naive Bayes) focused on the emails' body text. For the experiments, two datasets were used, one balanced and one imbalance, and the best performance, in terms of accuracy, was attained by the Random Forest classifier with 98.9% with Word2Vec on the balanced dataset.

Hybrid features involve combining multiple types of features to improve the performance of phishing email detection systems. Specifically, it refers to using features extracted through NLP methods from the body of the emails, along with features extracted from other parts of the email, such as URLs. By integrating these diverse features, the approach aims to enhance the accuracy and effectiveness of detecting phishing emails. Essentially, studies have argued that combining different types of information from email can lead to better detection results compared to using a single type of feature alone [2, 5, 13].

Bountakas et al. [13] propose a phishing email detection methodology, named HELPHED that focuses on the detection of phishing emails by combining Ensemble Learning methods with hybrid features. HELPHED Ensemble Learning can exploit all the information of the hybrid features by implementing two different algorithms to process the hybrid features separately in parallel methods consisting of two base learners to process the hybrid features separately, yet in parallel. The deployment of hybrid features provides a more thorough representation of emails. HELPHED accomplishes the best performance yielding a 99.42% F1-score, deployment of hybrid features provides a more thorough representation of emails.

However, NLP-based phishing email detection systems that rely on ML are primarily focused on the surface-level text of emails [14]. They analyze the specific words and phrases used, rather than understanding the deeper meaning or context of the text. As a result, these systems can struggle to detect phishing emails if the structure of a sentence is changed, synonyms are used, or other subtle modifications are made. Essentially, these systems lack the ability to fully grasp the semantics or the underlying intent behind the words, making them less effective at identifying cleverly disguised phishing attempts [14, 19].

Many surveys and studies have been conducted to examine the complex role of LLMs in cybersecurity. Due to the capabilities of LLMs in analyzing complex natural language patterns, studies are now able to explore a greater range of attack vectors in various contexts related to textual data [7,8]. The exploration of LMMs in cybersecurity is still in its early stages [21].

Najaf et al. [6] provide a comprehensive review of the current applications of LLMs in the field of both defensive and adversarial uses of LLMs and highlight how LLMs are used to enhance cybersecurity defenses by analyzing vast amounts of text data, including security logs. LLMs can identify emerging vulnerabilities and provide decision support for cybersecurity professionals. Although LLMs have demonstrated considerable promise in both protecting against and executing cyber threats, there is still a lot of work required to fully utilize their potential.

Another study conducted by Tanksale [9], explored the application of LLMs in identifying and mitigating cybersecurity threats, highlighting their potential to enhance threat-hunting. The research outlines the benefits of integrating LLMs in cybersecurity workflows while addressing challenges like bias, privacy, and computational efficiency. It suggests solutions such as training LLMs in cybersecurity-specific domains and incorporating contextual knowledge into threat-hunting methods.

Jiang [27] provides an overview of methodological steps for building an effective scam detector using LLMs, such as data collection, preprocessing, model selection, training, and integration into target systems. He also conducts a preliminary evaluation using GPT-3.5 and GPT-4 on an email, demonstrating their proficiency in identifying suspicious elements of phishing or scam emails.

Reference [28] presents the ChatSpamDetectors system that uses GPT-3.5 and GPT-4 to detect phishing emails, validates the result on a dataset, and receives 99.70% precision, recall, and accuracy on GPT-4. Reference [17] aims to evaluate the performance of ChatGPT-4 and Gemini Advanced on five assessment Indicators (Visual Cues, Social Proof, Appeal to Authority, Scarcity and Urgency, and Linguistic Cues). The study concludes that LLMs have significant potential to enhance cybersecurity measures against phishing attacks, but further research is needed to address their limitations and improve their real-world applicability. The reviewed studies on phishing email detection did not utilize large datasets (approximately less than 3000), which limited their ability to capture a broader range of phishing attempts and hallucinations in LLMs. Next, our study will introduce a primary methodology for integrating LLMs with existing email services to enhance phishing detection.

## III. METHODOLOGY

This section discusses the mythological steps to implement the phishing email detection framework that uses LLM to detect phishing emails. The general framework throughout includes dataset consideration, data processing, model section, generating prompts, and evaluation.

### A. Data Acquisition

Data selection is a crucial part of any machine learning and deep learning research. The diversity of the data can significantly influence the model's evaluation performance.

In this study, we evaluate four hybrid features for phishing detection:

*1)* **Email Body Content**: The email body content is the most causal aspect of this study. The email body contains the main content that the LLM can analyze using the social engineering aspect to detect whether an email is phishing or not.

*2)* **Email Subject Content**: Phishing emails frequently use urgent or enticing language in their subject lines to prompt immediate action from the recipient. These emails often contain requests for personal information or warnings about account security. Scammers typically use specific keywords that evoke fear, urgency, or excitement, such as "Immediate Action Required" or "You've Won." By identifying and compiling a list of these keywords, phishing detection systems can be improved to more accurately identify and flag potential phishing emails.

*3)* **Email Sender Information**: Suspicious sender addresses are crucial for identifying common characteristics of phishing, such as mimicking a famous brand but using a personal email address, providing insights into the patterns of email addresses used in phishing attempts.

*4)* **URLs**: Hyperlinks that are embedded within clickable elements like buttons or images. For example, a link might appear as <a href="phishing_website">Click here for more information</a>. By analyzing the domain name and path of the URL, it is possible to identify potential threats. If an email mentions a specific brand but the hyperlink does not match the brand's official domain, this discrepancy can indicate phishing activity. Essentially, checking the URL's authenticity can help detect phishing attempts.

The need for a diverse dataset to detect phishing emails is further emphasized by the alarming increase in phishing incidents and the sophisticated tactics employed by attackers [18]. This study utilized a publicly accessible dataset that is open to researchers. Various resources, such as GitHub and Kaggle, provide extensive data and tools for conducting such analyses, thereby facilitating further exploration and validation of the findings. Table 1 shows the dataset that is used in this study.

Phishing Pot [10, 28] is a GitHub repository that contains the latest phishing emails containing various brands and languages from contributors who upload real phishing emails that they received for researchers to study. This study includes 3000+ .eml files which represent an email message saved by an email application, such as Outlook. It contains the content of the message, along with the subject, sender, recipient(s), and date of the message.

With the evolution of LLMs can be used for offensive purposes, such as generating sophisticated phishing attacks and other cyber threats [9]. Cyber offenders can use the GenAI tools to develop phishing attacks [15, 20] called GenAI phishing. Our study will consider examining phishing emails generated by AI, and exploring how LLMs can detect these advanced techniques.

Additional datasets were accessed in various formats, specifically CSV and JSON. All datasets will be extracted to CSV files at the very end. JSON files can be converted to CSV files easily, while .eml files need to extract the subject and the sender from the header. Content from .eml files is in either text/plain or text/html type decoded with charset parameter

| Dataset | Legitimate | Phishing |
|---|---|---|
| Fraud Email Dataset [33] | 6742 | 5187 |
| Human-LLM-generated phishing-legitimate emails [34] | 726 (Human) + 998 (GPT) | 504 (Human) + 943 (GPT) |
| Phishing Email Data by Type [31] | 159 | 318 |
| Email Spam [32] | 3654 (non-spam) | - |
| Spear and traditional phishing [30] | - | 334 (Spear) + 3332 (Trad) |
| Phishing pot [10] | - | 3294 |

*Table 1. Details on the datasets.*

### B. Data Processing

Each dataset has its own unique structure, necessitating the development of a standardized data pipeline to ensure consistency across all datasets. The pipeline converts each dataset into a uniform format with two columns: "Email" and "Class."

- **Email**: This column will contain the entire email content, including the subject, sender, and body of the email. The format will be: SUBJECT: <subject>, FROM: <sender>, EMAIL: <email body>.
- **Class**: This column will categorize the emails into two standardized terms: "Phishing" (for fraudulent emails) and "Legit" (for legitimate emails).

In email communication, the incorporation of extensive HTML structures can enhance visual appeal. Therefore, the raw HTML content for emails is extracted into diverse text. From a content analysis perspective, many HTML tags are unnecessary and can be removed. These tags do not contribute to the actual content of the email. This study builds a custom HTML parser that extracts the HTML content into our optimized removal. This study retains only essential HTML elements, specifically the **<a>** and **** tags while focusing on pertinent attributes such as **href** and **src**. To mitigate long URL tokens, we limit the preservation of URL paths to the first ten tokens [28]. Additionally, it is noteworthy that some emails contain hidden text formatted with small font size (using **style="font-size:0px).** This hidden text often consists of randomly generated content intended to evade detection by users and security systems. These design choices warrant further exploration to understand their implications for email security and user awareness.

For handling email content encoded in Multipurpose Internet Mail Extension (MIME) base 64 format, a method for encoding binary data into text. There can be problems with decoding these emails properly due to issues with the encoding. This means that the email content might not be correctly converted back into its original format. The process involves first detecting if the email body is in base 64 format.

If it is, an attempt is made to decode it. If the decoding process fails due to minor issues, the email content is assigned as null. This means that instead of trying to work with improperly decoded content, the system will treat the email as having no content.

Data cleaning is an important part of the evaluation of the LLMs. After successfully extracting the email body content, the next step is to clean the data. Since LLMs have token limit rates, it's important to remove unnecessary tokens in the email. This includes:
- **Emojis and Special Characters**: These are removed as they are not essential for the analysis.
- **Whitespace Characters**: All occurrences of whitespace characters (such as carriage returns \r, newlines \n, tabs \t, and spaces \s) are replaced with a single space to standardize the text.

Any emails that are null (empty or invalid) and duplicate are removed from the dataset. Only emails with a length between 500 to 2000 characters (approximately 100 to 300 words) are selected for analysis. This ensures that the emails are of a manageable size for the LLMs. After cleaning, the final dataset consists of 6867 emails, with 4904 classified as phishing emails and 1962 as legitimate emails.

*C. Development of Evaluation Prompt*

Prompt engineering, which involves customizing input prompts, can enhance the accuracy of LLM responses. This study will apply the prompt engineering technique with role assignment, where the model is instructed to respond as a specific type of expert or in a particular style. The persona we give to the model is a cybersecurity expert who specializes in detecting phishing emails with a guide to the model on what aspect to look at to determine phishing email.

> "You are a cybersecurity expert specialized in detecting and analyzing phishing emails. Analyze the provided email (including subject line, body text, sender information, and links) to determine whether it is a phishing email or a legitimate email. Your result must follow the provided function call."

In response to a given prompt, the models will analyze the email based on the given prompt generate a response assign it to either the 'Phishing' or 'Legit' categories. Accompanying this classification is a phishing risk score, ranging from 'low', 'medium', or 'high', reflecting the likelihood that the email is a phishing attempt. Additionally, the framework elaborates on the rationale for its decision, offering a transparent view of its analytical process. The final output is delivered as a JSON response:
- "Is Phishing": A Boolean variable indicating whether the email is phishing or not.
- "Risk": where the LLMs indicated the certainty of this email being phishing.
- "Social Engineering Elements": A list of Social Engineering that the LLMs found in this email.
- "Actions": A list of recommended actions to apply.
- "Reason": A summary of the found from the LLMs on why this email is phishing (or legitimate).

```
{"Is_Phishing": {
   "type": "boolean",
   "description": "An email is phishing or not"
 },
 "Risk": {
   "type": "string",
   "description": "Categories as High, Medium, and Low"
 },
 "Social_Engineering_Elements": {
   "type": "List",
   "description": "A collection of social engineering elements from the email"
 },
 "Actions": {
   "type": "List",
   "description": "A collection of recommended action"
 },
 "Reason": {
   "type": "String",
   "description": "A brief reason why this email is phishing"}}
```

By combining structured outputs with the OpenAI inference, we ensure that the model's output exactly matches a specified JSON schema.

*D. Selection of Language Models*

This study will focus on examining the models that are the most widely used open-source models in current research: Llama-3.1-70b, gemma2-9b, Llama-3-8b, and Mistral-large-latest (123b) respectively. Additionally, Llama 3.1 was a newly released model at the time of this experiment, providing an opportunity to evaluate its performance in real-world scenarios. The selection of models with varying parameter sizes, from small to medium, was intentional for resource efficiency.

*E. Evaluation Metrics*

To facilitate a comparison several classification metrics are considered including accuracy, precision, recall, and F1-score metrics shown in Eq. 1, 2, 3, and 4, here the TP, TN, FP, and FN, denote true positive, true negative, false positive, and false negative values respectively.

$$\text{Accuracy} = \frac{TP + TN}{TP + TN + FP + FN} \quad (1)$$

$$\text{Precision} = \frac{TP}{TP + FP} \quad (2)$$

$$\text{Recall} = \frac{TP}{TP + FN} \quad (3)$$

$$F1 = 2 \times \frac{\text{Precision} \times \text{Recall}}{\text{Precision} + \text{Recall}} \quad (4)$$

## IV. RESULTS

The performance of all four LLMs can be seen in Table 2. All the LLMs achieved more than 80% accuracy in identifying phishing emails. With three models, Llama-3.1-70b stands out with the highest accuracy rate of 97.21%, closely followed by Gemma2-9b at 95.29% and Llama-3-8b at 92.39%, achieved accuracy more than 90% and Mistrial-large-latest 87.95%. Although all models have shown remarkable performance in accuracy, we need to consider other metric indicators to evaluate the overall performance. In terms of precision within the recall rate, Llama-3.1 has a high precision of 98.10% and a recall of 98.00%, indicating its effectiveness in accurately identifying and demonstrating its strong capability in detecting phishing emails. The low false positive rate of 4.7% suggests it is also good at avoiding legitimate emails being flagged as phishing, which can lead to reduced trust in the detection system.

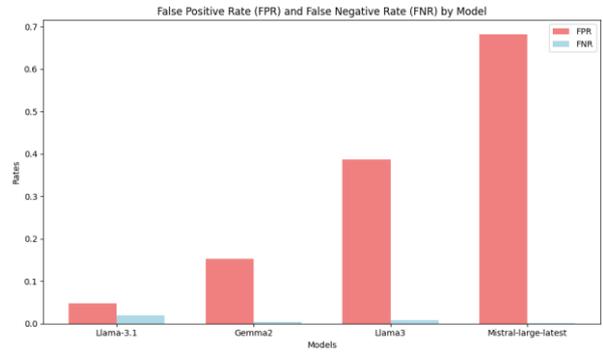

*Figure 1. False Positive Rate (FPR) and False Negative Rate (FNR) by Models. FPR on the left and FNR on the right. The lower the rate is the better the performance.*

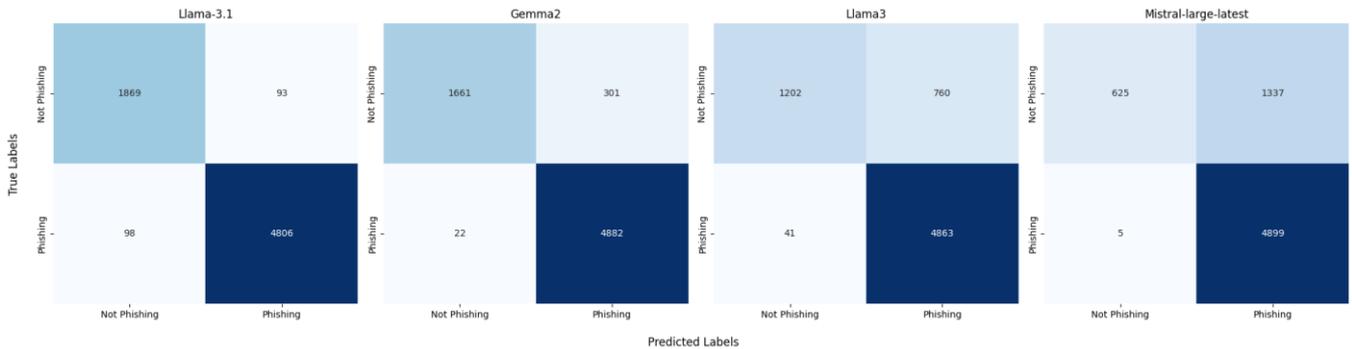

*Figure 2. Confusion matrix of the models. This image presents a comprehensive view of their performance in email classification.*

| Model | TP | FP | TN | FN | Precision | Recall | F1 | Accuracy |
|---|---|---|---|---|---|---|---|---|
| **Llama-3.1-70b** | 4806 | 93 | 1869 | 98 | **0.9810** | 0.9800 | **0.9805** | **0.9721** |
| **Gemma2-9b** | 4882 | 301 | 1661 | 22 | 0.9419 | 0.9955 | 0.9679 | 0.9529 |
| **Llama3-8b** | 4863 | 760 | 1202 | 41 | 0.8648 | 0.9916 | 0.8833 | 0.9239 |
| **Mistral-large-latest** | 4899 | 1337 | 625 | 5 | 0.7855 | **0.9989** | 0.8045 | 0.8795 |

*Table 2. Comparative Analysis of Performance Evaluation of LLMs on Accuracy, Precision, Recall, and F1.*

Gemma2 compared to Llama-3.1 indicates it may generate more false positives with a false positive rate (FPR) of 15% indicating it has a higher chance of predicting non-phishing (legit) emails as phishing. Observed the overall accuracy result with FPR and FNR, with the accuracy rate decreasing, the corresponding FPR increases and FNR decreases, generating a negative correlation between the accuracy rate and FPR and a positive correlation between FNR for the four models. The testing dataset used for evaluating the model is imbalanced, with 70% phishing emails and 30% legitimate emails. In such an imbalanced dataset, a model with a high FPR suggests that it's too aggressive in classifying emails as phishing. This aggressiveness might be due to the model's tendency to favor the majority class (phishing emails) rather than performing a thorough analysis of each email. As a result, Mistrial achieves a high recall rate, meaning it correctly identifies a large proportion of actual phishing emails; however, this comes at the cost of lower precision and F1 score. Lower precision indicates that a significant number of legitimate emails are incorrectly classified as phishing (false positives). The F1 score, which is the harmonic mean of precision and recall, is also lower, reflecting the trade-off between these two metrics. Essentially, while Mistrial is good at catching phishing emails, it also incorrectly flags many legitimate emails, reducing its overall effectiveness.

A reliability score quantifies how dependable an LLM is in generating correct, particularly in phishing email identification, where decision-making depends on the accuracy of the predictions. The reliability score is determined based on the predicted risk and the real class. With correct predictions, each LLM received one score on reliability. However, to increase the robustness of the models, we will assign half a mark if the predicated risk is medium, and the true class is phishing. Figure 3 shows the calculated reliability score for the LLMs over all 6867 emails.

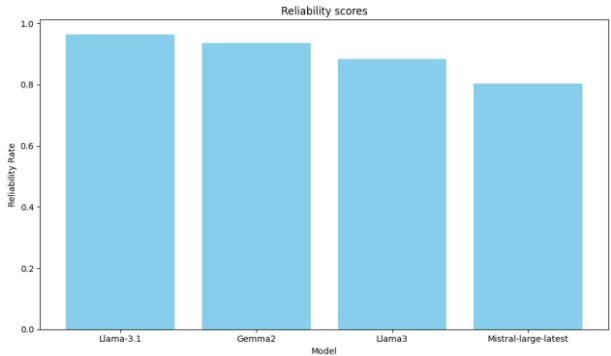

*Figure 3. The reliability score of models. This bar chart highlights the average reliability levels with each model on its classifications.*

As observed, Llama-3.1 and Gemma2 demonstrated significantly higher average reliability of 96.35% and 93.66%. The correlation between accuracy, FPR, and FNR can remain the same. Among the evaluated LLMs, the top performers in detecting phishing attempts were Llama 3.1 and Gemma2 followed by Llama3 and Mistrial are less effective.

## V. Discussion

### A. LLM Capabilities for Phishing Email Detection

In this section, we further analyze the detection capabilities of phishing email results generated by LLMs. Figure 4 presents an example of a phishing email sourced from our dataset. To safeguard user privacy, all sensitive information has been replaced with the placeholder "phishing@pot" [10].

This email masquerades as a system notification from Facebook, informing the user that their account has been accessed from a new device, and requests verification of the user's involvement in this action. At first glance, the visual structure of the email does not provide clear indications of its phishing nature. However, a closer examination reveals critical discrepancies: the sender's email address does not originate from the domain facebook.com, and the link embedded in the "Report the user" button redirects to a fraudulent website.

All LLMs are capable of accurately identifying suspicious elements in this email. A report from Llama 3.1 indicates that the email address '5a83h@92e4fsmb2e.com' does not match the displayed sender, which is denoted as 'Facebook.' Furthermore, the email contains potentially harmful links: 'ssecnewsso' is present in both the 'Report the user' and 'Yes me' buttons. Additionally, the email includes a hidden image with a suspicious URL, http://thema214.com/track/o49, which may be intended for tracking purposes.

Gemma2 identifies several linguistic indicators that suggest potential phishing activity. The first element is the use of the recipient's username in the greeting, as exemplified by the phrase "Username in Greeting." The rationale for this observation is that it is atypical for official correspondence to include the recipient's username. Additionally, the email employs generic greetings, which lack personalization, such as "Hi phishing@pot." Furthermore, there is an element of urgency conveyed by the statement that someone attempted to log into the account. Finally, the email exhibits poor grammar and inconsistent formatting, which serves to undermine its legitimacy.

### B. Limitation in Detection

This section addresses the key issues identified in the overall study. In Section IV, we examine the performance in terms of FPR, focusing on emails misclassified by LLMs. Through analysis supplemented by human feedback, we observed the limitations in the dataset's representation. Specifically, we identified cases where emails were labeled as phishing despite lacking characteristics typically associated with phishing attempts, such as no requests for any personal information or urgency cues, suggesting that the data may have been incorrectly annotated. Another hypothesis that emerged regarding this situation is that certain emails might lack phishing indicators, such as urgent content, but could include malicious URLs not captured due to the conversion of raw email data into plain text in the original dataset. This highlights the potential loss of critical features during preprocessing, which warrants further investigation.

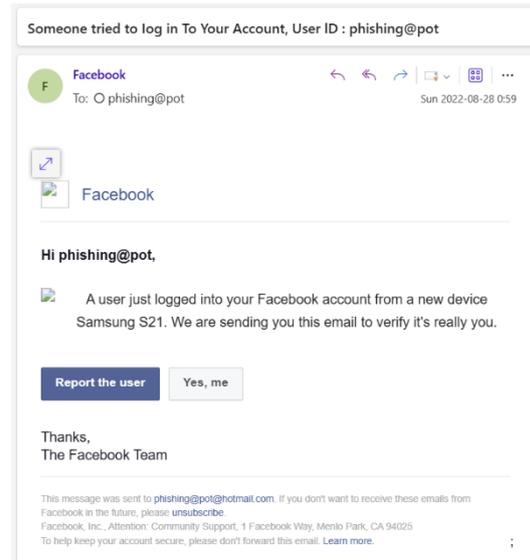

*Figure 4. An example of a phishing email from the dataset.*

For actual misclassified emails, we observed that LLM misclassified phishing emails that show personalization such as addressing the recipient by name, referencing their current position and company, and having a clear and relevant purpose. This level of personalization makes the LLM consider emails to appear legitimate. Additionally, the tone and language used are professional and consistent with a genuine offer. However, this type of email is spearing phishing that targets specific individuals to steal sensitive information.

We can observe the same limitations when solely analyzing URL-based features for phishing detection. LLMs might incorrectly flag some legitimate emails that contain uncommon URL domains as phishing emails. LLMs might misinterpret these materials, raising concerns that they could be phishing attempts to compromise personal information. Additionally, we observed that phishing emails containing shortened URLs can bypass LLM detection, indicating that phishers can use strategies to evade URL detection by encoding fraudulent website addresses with hexadecimal character codes [4]. For example, a link in a phishing email like http://bit.ly/fdasgfcxv might appear legitimate to LLM because it uses a well-known URL shortener (bit.ly), but the actual destination (fdasgfcxv) could be a phishing website. This technique makes it harder for LLM to identify phishing attempts.

*C. Integration with Existing Systems*

Integrating LLMs with existing services represents a transformative approach to enhancing security functionality and efficiency in current cybersecurity. To automatically detect phishing emails with existing email services, we utilized automated routine tasks like email analysis, threat detection, and report generation using the LLM.

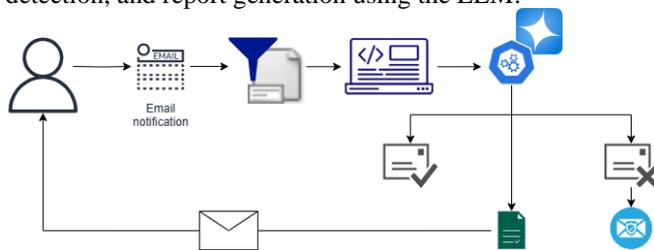

*Figure 5. An overall framework flow of automated routine tasks for phishing detention on regular email services.*

Figure 5 presents an overview of the automated routine framework's architecture that harnesses the power of LLMs. The framework can be divided into the following steps:

*1)* **New Email Received**: The routine will be triggered when a new email is received in the user mailbox.

*2)* **Filter emails that come from a Trusted source**: The system task will first filter emails based on the sender's information such as emails from a legitimate office address and email addresses that are on whitelisting.

*3)* **HTML parsing**: Retrieve the MIME type of subject, from, and the "text/html" body of the email. Applied the data cleaning processing proposed in section III part B for HTML parsing.

*4)* **Run OpenAI API request**: After processing, apply the prompt response request with the setting defined in section III part C on Llama3.1-70b.

*5)* **Email classification result**: Once the email is identified as phishing, report this email as phishing to the email system, move the email to the spam folder, and send the report result generated by the LLM to the user explaining why this email is phishing and corresponding actions to apply. The following is an example summary report generation using the LLM on the phishing email shown in figure 4.

> This email displays multiple red flags that indicate it is a potential phishing attempt. The sender's email address is suspicious, the greeting uses the recipient's username, and the email contains urgent calls to action and clickable links that are likely malicious. It is important to exercise caution and follow the recommended actions to protect yourself from account compromise.
> Do not interact with any links or buttons in the email. Verify account activity through a secure login on the official Facebook website. Report the email to Facebook's support team.

The integration of LLMs in email services enhances the identification of emails that were not flagged by existing systems. This technology also aids in recognizing new phishing and social engineering tactics that are currently being addressed in defense systems. The Microsoft Defender team is beginning to develop LLMs to improve threat classification, which helps keep malicious emails out of users' inboxes. Additionally, it provides Security Operations (SecOps) teams with better insights into attacker techniques [35]. As industries increasingly explore the potential of LLMs, it becomes critical to understand their implementation within established systems to maximize benefits while addressing phishing challenges.

*D. Feature improvements*

In part B, we reviewed and discussed some limitations of our current studies by analyzing the false positives. However, these false positives also point out the weakness of LLMs, which attackers can intentionally exploit. This section will discuss some potential improvements that can be made to the feature studies.

LLMs can be vulnerable to adversarial attacks where malicious actors craft emails specifically designed to bypass detection [36]. Research on improving the robustness of LLMs against phishing attacks is still in its early stages. Robustness against diver Attacks can be achieved with fine-tuning training the LLM to recognize and learn from a wide range of examples of comprehensive phishing attack types [37].

Our diverse dataset lacked some phishing email characteristics, leading to potential misinterpretations. Al-Subaiey et al. [38] created a comprehensive dataset by combining six widely used spam email datasets, carefully selected based on their unique attributes from various sources, to create a comprehensive resource for analysis. This dataset is notable for offering one of the largest collections of approximately 82,500 emails, providing a rich foundation for feature research in phishing email detection.

LLMs are not specifically designed to identify URLs or domain names. As a result, they might mistakenly flag URLs from uncommon websites as suspicious or phishing attempts. To address this issue, the proposed solution involves adding a new URL analyzer layer to the system architecture. This layer will extract all URLs from the email and then use an API endpoint security tool (virustotal.com) to generate a security report on each URL. This report will assist the LLM in making more accurate decisions about whether an email is a phishing attempt. Essentially, this approach aims to improve the LLM's ability to correctly identify phishing emails by providing additional context and analysis for URLs.

## VI. Concussion

This study proposed and studied an LLM-based phishing email detection system. The results showed that our system using Llama-3.1-70b achieved an accuracy of 97.21%, outperforming other models. Through detailed analysis of LLM performance and responses, we can confirm the ability of LLMs to extract key hybrid features in emails, prioritize them, and generate accurate responses, assisting the effectiveness of the defense team in phishing detection. However, we also address several challenges associated with using LLMs in phishing detection and propose potential solutions to these challenges, such as improving LLM robustness with phishing email-specific data and incorporating an extra URL detection layer.

The rapid development of LLMs with greater performance capabilities offers new opportunities for in-depth research in this field. These advanced models can potentially provide more effective tools and techniques for identifying and mitigating phishing threats, thereby enhancing cybersecurity measures.